\documentstyle[preprint,aps,epsfig]{revtex}
\begin{document}
\draft
\title{Dynamic electrorheological effects and interparticle force \\ 
 between a pair of rotating spheres}
\author{Jones T. K. Wan$^{1}$, K. W. Yu$^{1}$ and G. Q. Gu$^{1,2}$}
\address{$^{1}$Department of Physics, The Chinese University of Hong Kong,\\
         Shatin, New Territories, Hong Kong, China \\
 $^{2}$College of Computer Engineering, University of Shanghai for Science 
 and Technology, \\ Shanghai 200 093, China}
\maketitle

\begin{abstract}
We consider a two-particle system in which a particle is held fixed, 
and the other one rotates around the axis perpendicular to the
line joining the particles centers. The rotating particle leads to a
displacement of its polarization charge on the surface.
Our results show that the rotational motion of the particles generally 
reduces the force between the particles.
The dependence of interparticle force on the angular velocity of rotation
will be discussed.
\end{abstract}
\vskip 5mm

\section{Introduction}

The prediction of the strength of the electrorheological (ER) effect 
is the main concern in the theoretical investigation of ER fluids [1--5].
An ER fluid is 
a suspension of particles with high dielectric constant (conductivity) 
in a host medium of low dielectric constant (conductivity). 
The ER effect originates from the induced interaction between the 
polarized particles in an ER fluid. 
Upon the application of an external electric field, polarization charges
are induced on the particle surfaces, leading to anisotropic forces
between the particles. Under the influence of the ER effect, the particles 
in an ER fluid initially aggregate rapidly into chains within milliseconds
and then the chains aggregate into columns within seconds [3,4]. 
The rapid field-induced transition between the fluid and solid phase 
makes this material both important for wide industrial applications 
such as shock absorbers, dampers and clutches, 
as well as for experimental and theoretical investigations.

In deriving the induced forces between the particles, existing theories 
assume that the particles are at rest [6--10]. In a realistic situation, 
the fluid flow exerts force and torque on the particles,
setting the particles in both translational and rotational motions.
For instance, the shear flow in an ER suspension exerts a toque on the 
particles, which leads to the rotation of particles about their centers [11].
Recent experiments showed that the induced forces between the rotating 
particles are markedly different from the values predicted by the 
existing theories which have not taken the motion of particles into 
account [12].

Wang designed a delicate experiment, in which a spherical particle 
is held fixed and the other one rotates uniformly, to investigate 
the dynamic ER effects \cite{Wang}. 
The experimental result is that the force exerted on the rotating particle 
is smaller than that on the same particle at rest. 
In order to explain this effect, 
we formulate a model, which describes the relaxation process of the 
polarized charge on the surfaces of the rotating particles.
We will compute the force between the polarized particles by the multiple
image method \cite{Wan}. We will show that the induced 
force is inversely proportional to the square of the product of the 
relaxation time and the angular velocity of the rotating particle.

\section{Dipole moment of a rotating dielectric sphere}

In this section, we consider two spherical particles, one of which is 
held fixed, and the other one rotates around the axis perpendicular to the
line joining the particles centers (Fig.\ref{angular-dependent}).
Both particles are dielectric spheres of dielectric constant $\epsilon_1$, 
suspended in a host medium of dielectric constant $\epsilon_m$.
The rotating (fixed) sphere has a radius $a$ ($b$).
In this section, we neglect the mutual polarization between the particles. 
This assumption is valid if the distance between the particles is large.
Upon the application of an electric field, polarization charges are induced
on the surface of the spheres and there are induced dipole moments.
Suppose the applied field is given by $\vec{E}_0$=$E_0 \hat{z}$,
the induced dipole moment is given by $\vec{p}_{a0}$=$p_{a0}\hat{z}$, 
where
\begin{equation}
p_{a0}=\epsilon_m E_0 \left({\epsilon_1-\epsilon_m \over 
 \epsilon_1+2\epsilon_m} \right) a^3, 
\end{equation}%1
and a similar expression for the fixed sphere.
In the presence of a rotational motion, a displacement of polarization 
charge occurs on the particle surface.

Suppose the sphere is undergoing an anti-clockwise rotation with an 
angular velocity $\vec{\omega}=-\omega \hat{y}$.
The induced dipole moment will rotate with the same angular velocity, 
and the rate of change of the induced dipole moment is given by:
\begin{equation}
{d \vec{p}\over dt} = \vec{\omega}\times\vec{p}.
\end{equation}%2
However, the induced dipole moment can relax back to its original 
orientation (along $\vec{E}_0$) and magnitude.
If the relaxation process is characterized by a relaxation time 
$\tau_r$, we have: 
\begin{equation}
{d \vec{p}\over dt} = -{1\over\tau_r}(\vec{p}-\vec{p}_{a0}).
\end{equation}%3
When a steady state is reached, we have:
\begin{equation}
\vec{\omega}\times\vec{p}-{1\over\tau_r}(\vec{p}-\vec{p}_{a0})=0.
\end{equation}%4
Solving for the above equation, we obtain:
\begin{equation}
\vec{p} = {\vec{p}_{a0} + \tau_r(\vec{\omega}\times\vec{p}_{a0})\over 
 1+(\omega\tau_r)^2}. \label{rotating-dipole}
\end{equation}%5
The result of Eq.(\ref{rotating-dipole}) has a simple geometric 
interpretation. It can be shown that the angle between
$\vec{p}_{a0}$ and $\vec{p}$ is equal to $\phi$ and the magnitude of 
$\vec{p}$ is given by
$|\vec{p}|={p}_{a0}\cos{\phi}$, where $\phi$ is given by 
$\tan{\phi}=\omega\tau_r$.

\section{Multiple image dipoles for a pair of dielectric spheres}

In this section, we generalize the multiple image method \cite{Wan} to 
calculate the force between the polarized spheres, when the induced 
dipole moments are not parallel.
Consider two dielectric spheres, the radii of which are given by $a$ 
and $b$ respectively and the separation between their center is $r$.
Suppose both the spheres are polarized with induced dipole moments 
$\vec{p}_{a0}$ and $\vec{p}_{b0}$ respectively.
Due to mutual polarization effect, the dipole moment of each sphere will 
induce an image dipole on the other sphere.
The induced image dipole will further induce another image dipole on the 
original sphere and hence multiple images are formed \cite{Wan}.
To express the image dipole moment, it is convenient to separate the 
original dipole moments into $x$ (perpendicular to $\vec{E}_0$) 
and $z$ components (parallel to $\vec{E}_0$): 
\begin{equation}
\vec{p}_{a0}=p_{a0x}\hat{x} +p_{a0z}\hat{z}, 
\quad\vec{p}_{b0}=p_{b0x}\hat{x} +p_{b0z}\hat{z}.
\end{equation}%6
The total dipole moments are given by:
\begin{equation}
\vec{p}_{a}=p_{ax}\hat{x} +p_{az}\hat{z},
\quad\vec{p}_{b}=p_{bx}\hat{x} +p_{bz}\hat{z}. 
\end{equation}%7
The $x$-components can be found by using the multiple image method 
\cite{Wan}:
\begin{eqnarray}
p_{ax} &=& \sum_{n=1}^\infty \left[
 {p_{a0x}(b \sinh \alpha)^3 (-\tau)^{2n-2} \over (b \sinh n\alpha + a 
\sinh (n-1)\alpha)^3 }
  +{p_{b0x}(a \sinh \alpha)^3 (-\tau)^{2n-1} \over (r \sinh n\alpha)^3} 
  \right],
\label{pax}\\ 
p_{bx} &=& \sum_{n=1}^\infty \left[
 {p_{b0x}(a \sinh \alpha)^3 (-\tau)^{2n-2} \over (a \sinh n\alpha + b 
\sinh (n-1)\alpha)^3 }
  +{p_{a0x}(b \sinh \alpha)^3 (-\tau)^{2n-1} \over (r \sinh n\alpha)^3} 
  \right].
\label{pbx} 
\end{eqnarray}%8--9
The $z$-components can be found similarly \cite{Wan}:
\begin{eqnarray}
p_{az} &=& \sum_{n=1}^\infty \left[
 {p_{a0z}(b \sinh \alpha)^3 (2\tau)^{2n-2} \over (b \sinh n\alpha + a 
\sinh (n-1)\alpha)^3 }
  +{p_{b0z}(a \sinh \alpha)^3 (2\tau)^{2n-1} \over (r \sinh n\alpha)^3} 
  \right],
\label{paz}\\ 
p_{bz} &=& \sum_{n=1}^\infty \left[
 {p_{b0z}(a \sinh \alpha)^3 (2\tau)^{2n-2} \over (a \sinh n\alpha + b 
\sinh (n-1)\alpha)^3 }
  +{p_{a0z}(b \sinh \alpha)^3 (2\tau)^{2n-1} \over (r \sinh n\alpha)^3} 
  \right].
\label{pbz} 
\end{eqnarray}%10--11
The factor $(-\tau)$ and $(2\tau)$ apply to the transverse 
and longitudinal image dipole moments respectively.
The factor $\tau$ is known as the dielectric contrast and is given by:
\begin{equation}
\tau={\epsilon_1-\epsilon_m \over \epsilon_1+\epsilon_m}.
\end{equation}%12
The parameter $\alpha$ is given by:
\begin{equation}
\cosh\alpha={r^2 -a^2-b^2\over 2 a b}.
\end{equation}%13
We should remark that the present generalization is an approximation only,
because there is a more complicated image method for a dielectric sphere 
\cite{Poladian}.
However, in the limit $\tau \to 1$, the above expressions reduce to the
results of perfectly conducting spheres. We expect that this approximation 
will be good at high contrast ($\tau \to 1$). 
We have checked the validity of the approximation by comparing the above 
analytic expressions with the numerical solution 
of an integral equation method \cite{Yu99}.
The force between the spheres is given by the following expression, 
based on the energy considerations \cite{Yu,Jackson}:
\begin{equation}
\vec{F}={1\over 2}\nabla \left[ \vec{E}_0\cdot(\vec{p}_a+\vec{p}_b) \right]
=F_r\hat{r}+F_\theta\hat{\theta}.
\end{equation}%14
It should be remarked that $\vec{F}(\theta)=\vec{F}(\theta+\pi)$ 
by examining Fig.\ref{angular-dependent}. 
We will calculate the results for an anti-clockwise rotation 
($\vec{\omega}=-\omega\hat{y}$) only.
The results for a clockwise rotation can be found similarly.

\section{Numerical results}

In this section, we report the separation and 
angular dependence of the force between the spheres.
We will calculate the radial component ($F_r$) and the tangential
component ($F_\theta$) of the force and discuss their dependence on 
$\omega\tau_r$. Moreover, we let $b=a$ for convenience.

In Fig.\ref{force-distance-dependent}, we plot the radial force between 
the spheres against the separation parameter $\sigma$, defined by
$\sigma=r/(a+b)$. 
Two cases are considered, namely, the transverse and longitudinal field 
cases. In the transverse (longitudinal) field case, the line joining the 
spheres is perpendicular (parallel) to the applied field.
We first examine the transverse field case, with 
$\omega\tau_r$ being chosen to be 0.1, 1 and 10 respectively.
The dielectric contrast is chosen to be $\tau=1/3$ (low contrast)
and $\tau=9/11$ (high contrast).
From Fig.\ref{force-distance-dependent}, our results show that the 
rotational motion of the particle generally reduces 
the force between the particles. In order to see the effects more
clearly, we plot in Fig.\ref{force4-distance-dependent} the products 
$F_T \sigma^4$ and $F_L\sigma^4$ 
against $\sigma$. It can be seen that for a large separation $(\sigma>3)$,
these quantities tend to constant, indicating that
the force varies as $\sigma^{-4}$. For a small separation $(\sigma<1.5)$, 
the magnitude of the transverse force increases rapidly.
The longitudinal field case shows a similar behavior: the rotational motion 
generally reduces the magnitude of the interparticle force.

We next discuss the angular dependence of the interparticle force 
(Fig.\ref{force-angular-dependent}).
We concentrate on the radial component $(F_r)$ first. We can see 
that, for a small angular velocity
$(\omega\tau_r =0.1)$, the radial component attains its minimum and 
maximum value at $\theta\approx 0$
and $\theta\approx\pi/2$ respectively, which correspond to the 
longitudinal and transverse field case
(Fig.\ref{force-distance-dependent}). 
For the tangential force, $F_\theta\approx 0$ at
$\theta=0$ and $\theta=\pi/2$. 
This is expected because when the angular velocity of the particle
is small, the effect of rotation can be neglected.
However, when the sphere rotates faster $(\omega\tau_r =1)$, 
the peak shifts towards
the direction of smaller $\theta$. Similar conclusion can be drawn for 
the tangential force.
In the lower panels of Fig.\ref{force-angular-dependent}, the tangential 
force no longer vanishes at $\theta=0$ and $\theta=\pi/2$ 
for $\omega\tau_r = 1$. 
From the above results, we conclude that the nonrotating particle
is forced to move in the direction of increasing $\theta$. 
More precisely, if originally the two particles are at
$\theta=\pi/2$ while there is no rotational motion, the particles will 
repel each other and there is no tangential motion.
However, if one of the particle begins to rotate, a tangential force 
will act on the other particle, forcing it to
move downwards, since $F_\theta>0$ at $\theta=\pi/2$.

In the opposite limit, when the angular velocity is large 
($\omega\tau_r=10$), the peak of the radial force
will return to the original position ($\theta=0$ and $\theta=\pi/2$), 
and $F_\theta = 0$
when $\theta=0$ or $\theta=\pi/2$. Furthermore, the magnitude of the 
force is reduced to nearly one half of its original value
(Fig.\ref{force-mag-angular-dependent}). 
It is because when the angular velocity is large, from  
$p'_{a0}=p_{a0}\cos{\phi}$ and $\tan{\phi}=\omega\tau_r$
(Fig.\ref{angular-dependent}), the magnitude of the steady-state dipole 
moment will be small. 
The situation is reminiscent of a polarized sphere interacting with a 
dielectric sphere with a vanishing polarization charge. 

\section{Discussion and Conclusion}

Here a few comments on our results are in order.
In this work, only one of the particles has been assumed to rotate.
In a realistic situation, all the particles can rotate. 
As our theory is general, it can readily be applied to the case when all
the particles are in rotational motion.
 
So far, our proposed relaxation time has no microscopic origin.
If the relaxation process is originated from a finite conductivity of 
the particle or host medium, then we can estimate the relaxation time 
based on the Maxwell-Wagner theory of leaky dielectrics \cite{MW}.
For a (nonrotating) spherical inclusion embedded in a host medium, 
the expression is 
$$
\tau_r=(\epsilon_1 + 2 \epsilon_m)/(\sigma_1 + 2 \sigma_m),
$$ 
where $\sigma_1, \sigma_m$ denote the conductivity of the sphere and 
host medium respectively.

We may extend the Maxwell-Wagner theory to polarization relaxation 
of rotating particles. In this case, we should
add a term $\rho_P \vec{v}$ to the polarization current density, where 
$\rho_P$ is the polarization charge density and 
$\vec{v}=\vec{\omega} \times \vec{r}$ is the rotating velocity. 
However, it is not possible to convert the extra term into a dielectric 
constant and the generalization becomes more complicated due to 
the nonuniform polarization charge density inside the rotating 
spherical inclusions. We are currently examining the solution of the 
more complicated boundary-value problem.

However, we can make a crude estimate of the relaxation time. 
According to the geometry of the rotating sphere, the relaxation of 
polarization charge can be along the particle-fluid interface, so
$$
1/\tau_r=\sigma_1/\epsilon_1 + \sigma_m/\epsilon_m.
$$
When $\epsilon_1$ is large compared with $\epsilon_m$ and $\sigma_m=0$ 
(insulating host medium), both the expressions for $\tau_r$ reduce to 
the same result.

In conclusion, we have developed a model to calculate the interparticle 
force between a pair of dielectric spheres, in which one sphere is held 
fixed and the other is in rotational motion.
We have calculated the force between the spheres by using the multiple 
image method. We have found that
the effect of rotation is large when the angular velocity of the sphere 
is large ($\omega \tau_r >> 1$).
To investigate the dynamic ER effects, we have solved a differential 
equation which describes the dynamic polarization of the rotating particles.
The theoretical problem described here can be realized in 
experiments \cite{Wang}.

\section*{Acknowledgments}

This work was supported by the Research Grants Council of the Hong Kong 
SAR Government under project numbers CUHK 4290/98P and CUHK 4284/00P. 
G. Q. G. acknowledges financial support from the Key Project of the 
National Natural Science Foundation of China, under grant 19834070.
G. Q. G. and K. W. Y. are grateful to Dr. Z. W. Wang, who informed us 
his experiment on the dynamic ER effects of rotating particles and 
for many fruitful discussions.

\begin{figure}[h]
\caption{A rotating polarized dielectric sphere interacting with another 
non-rotating polarized dielectric sphere.}
\label{angular-dependent}
\end{figure}

\begin{figure}[h]
\caption{Interparticle force $F_T$ and $F_L$ plotted against $\sigma$.
The subscript $T$ denote the transverse dipole $(\theta=\pi/2)$ and $L$ 
the longitudinal dipole $(\theta=0)$. The rotational 
motion of the particle generally reduces the interparticle force.}
\label{force-distance-dependent}
\end{figure}

\begin{figure}[h]
\caption{Similar to Fig.\ref{force-distance-dependent},
but for $F_T \sigma^4$ and $F_L \sigma^4$. The magnitude
of the forces increases rapidly when the separation becomes small.}
\label{force4-distance-dependent}
\end{figure}

\begin{figure}[h]
\caption{The angular dependence of force. The angle is expressed in unit 
of $\pi$, and the separation parameter is taken to be $\sigma$=1.1. 
The peaks shift to the direction of smaller $\theta$. 
For $\omega\tau_r =1$, the angular locations 
$\theta=0$ and $\theta=\pi/2$ are no longer the equilibrium position 
of the tangential force.} 
\label{force-angular-dependent}
\end{figure}

\begin{figure}[h]
\caption{The angular dependence of the magnitude of the force. 
The magnitude of the force is reduced to nearly one half of the 
original value.}
\label{force-mag-angular-dependent}
\end{figure}

\newpage
\centerline{\epsfig{file=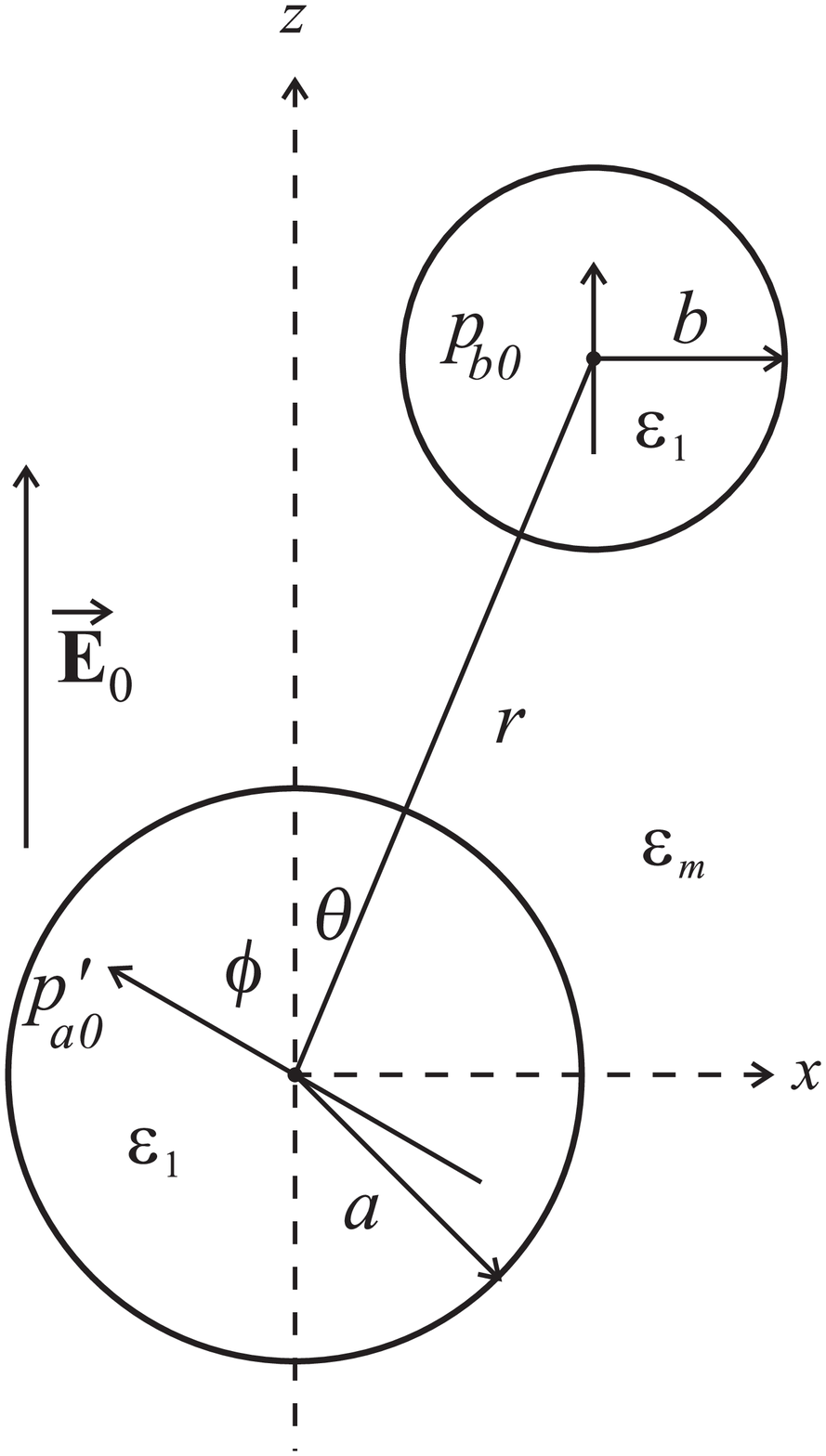,height=15 cm}}
\centerline{Fig.1/Wan/PRE-EP7288}

\newpage
\centerline{\epsfig{file=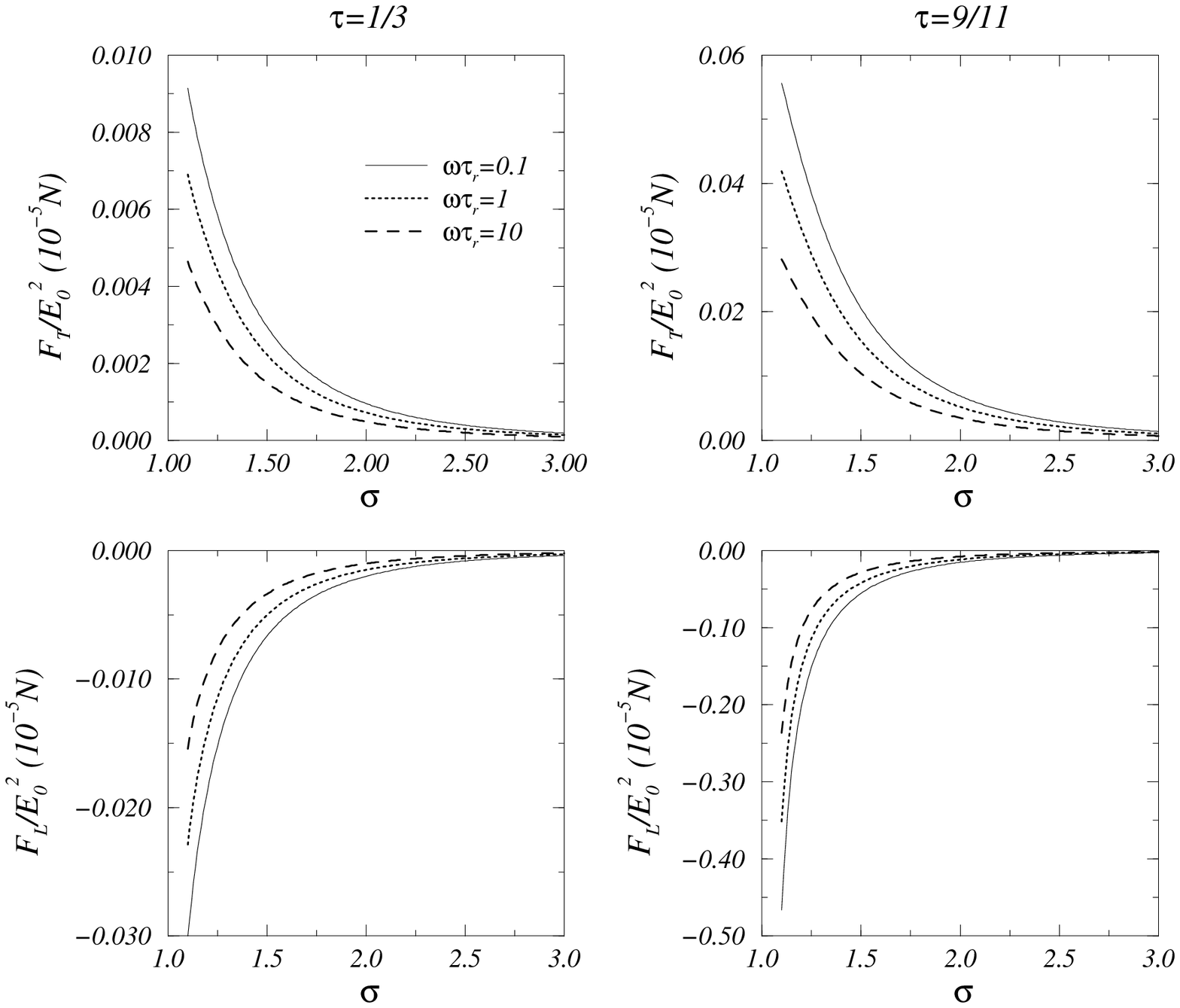,width=\linewidth}}
\centerline{Fig.2/Wan/PRE-EP7288}

\newpage
\centerline{\epsfig{file=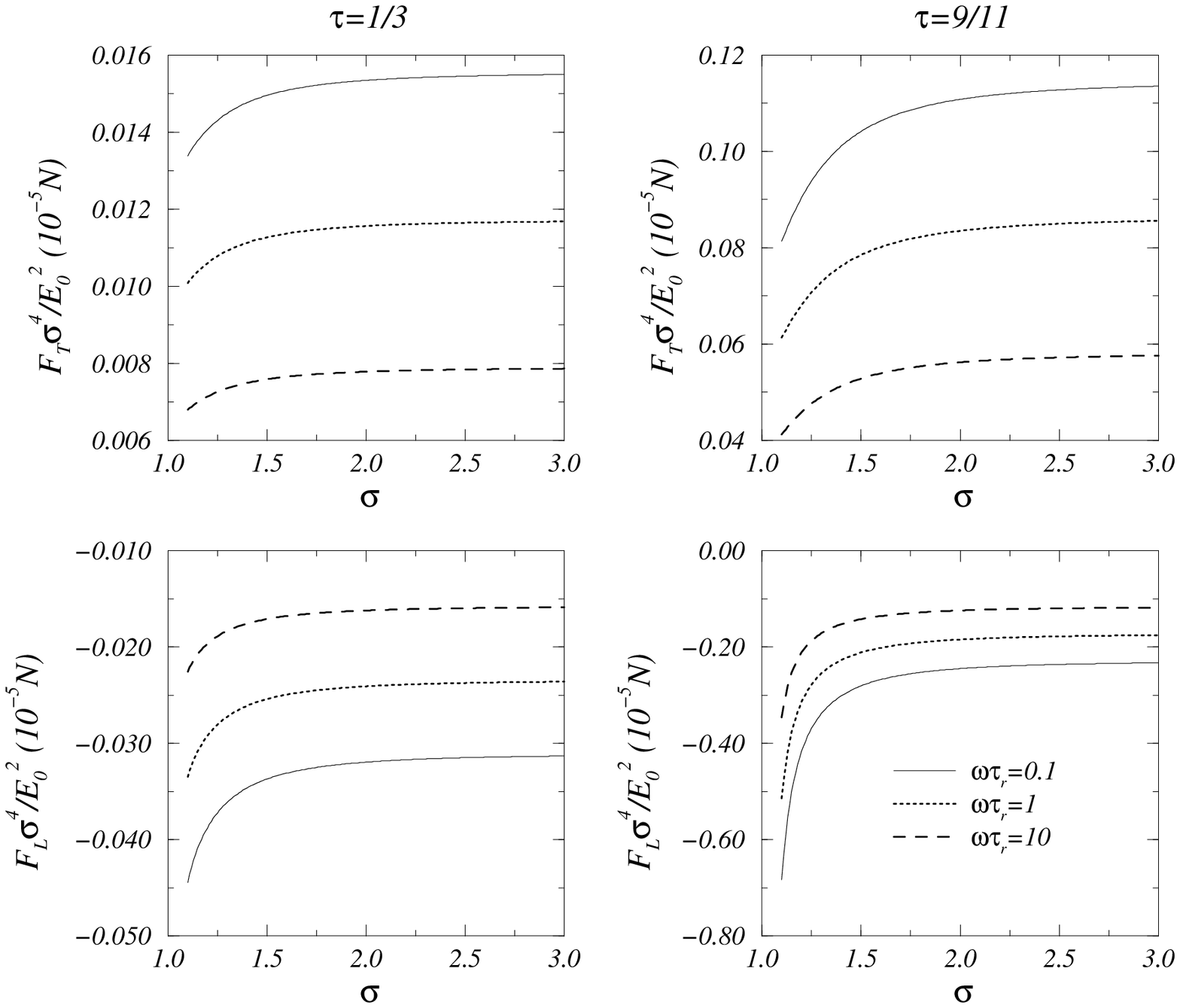,width=\linewidth}}
\centerline{Fig.3/Wan/PRE-EP7288}

\newpage
\centerline{\epsfig{file=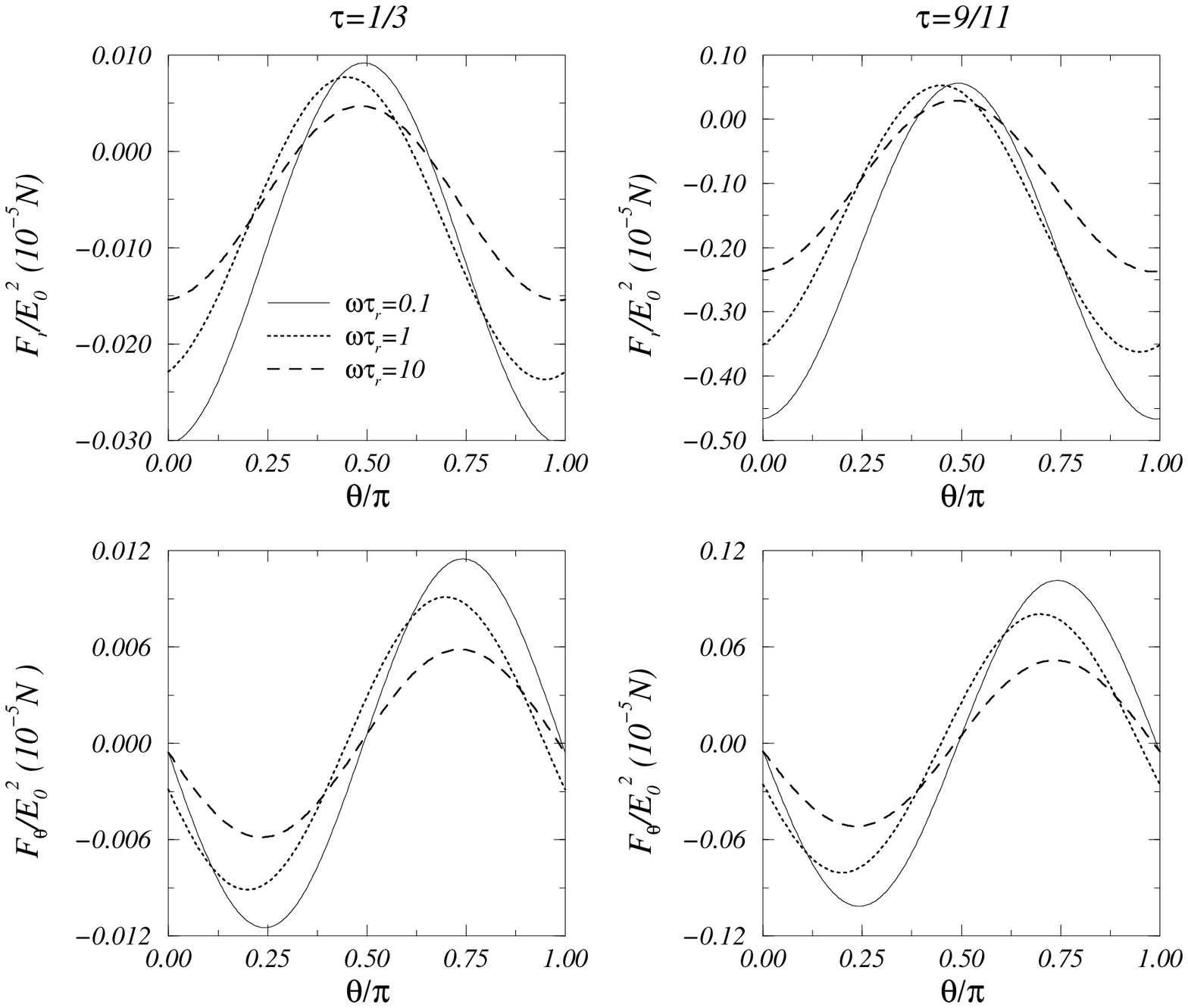,width=\linewidth}}
\centerline{Fig.4/Wan/PRE-EP7288}

\newpage
\centerline{\epsfig{file=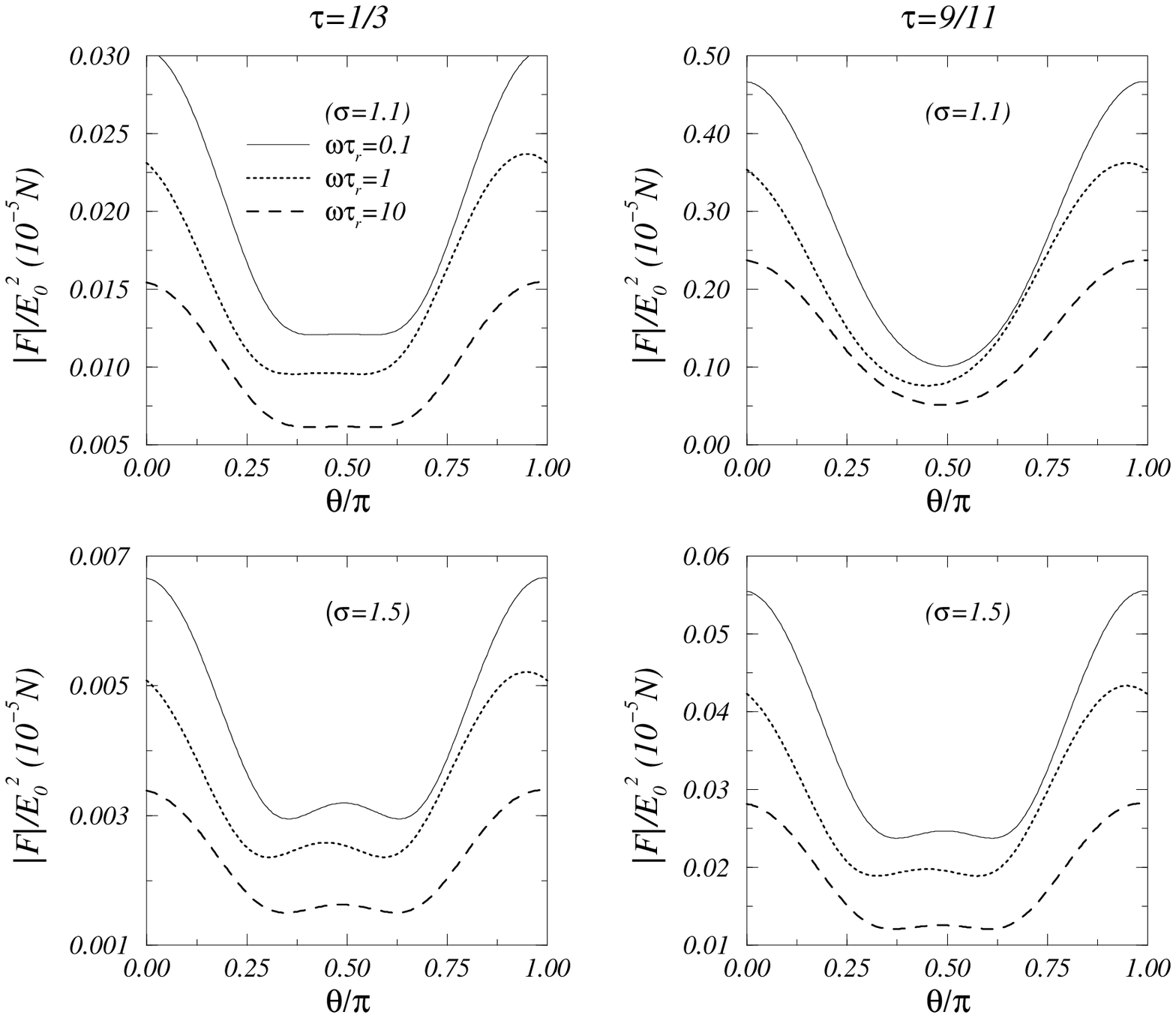,width=\linewidth}}
\centerline{Fig.5/Wan/PRE-EP7288}

\end{document}